\documentclass[slac_one]{revtex4}
\usepackage{graphicx}
\usepackage{fancyhdr}
\begin{document}

%%pbr definitions

\newcommand {\thw}        {\theta_{\mathrm{W}}}
\newcommand {\ntau}        {\nu_{\tau}} 
\newcommand {\MZ}      {M_{\rm{Z}}}
\newcommand {\MW}      {m_{\mathrm{W}}}
\newcommand {\MH}      {m_{\rm{H}}}
\newcommand {\Mt}      {m_{\rm{t}}}
\newcommand {\Mb}      {m_{\mathrm{b}}}
\newcommand {\GZ}      {\Gamma_{\mathrm{Z}}}
\newcommand {\GW}      {\Gamma_{\mathrm{W}}}
\newcommand {\Ghad}    {\Gamma_{\mathrm{had}}}
\newcommand {\Gb}      {\Gamma_{\mathrm{b}}}
\newcommand {\Gc}      {\Gamma_{\mathrm{c}}}
\newcommand {\alphamz}    {\alpha(\MZ)}
\newcommand {\alphasmz}    {\alpha_{\rm{s}}{\rm(\MZ)}}
\newcommand {\chisq}       {\chi^{2}}

\newcommand  {\Zzero}   {\mbox{${\mathrm{Z}}$}}
\newcommand {\ff}         {{\rm f}\overline{\rm f}}
\newcommand {\gVf}         {g_{\rm{Vf}}}
\newcommand {\gAf}         {g_{\rm{Af}}}
\newcommand {\gVl}         {g_{\rm{Vl}}}
\newcommand {\gAl}         {g_{\rm{Al}}}
\newcommand {\gVe}         {g_{\rm{Ve}}}
\newcommand {\gAe}         {g_{\rm{Ae}}}
\newcommand {\gVm}         {g_{\rm{V}\mu}}
\newcommand {\gAm}         {g_{\rm{A}\mu}}
\newcommand {\gVt}         {g_{\rm{V}\tau}}
\newcommand {\gAt}         {g_{\rm{A}\tau}}
\newcommand {\ALR}         {A_{\rm {LR}}}
\newcommand {\swsq}       {1-\MW^2/\MZ^2}
\newcommand {\swsqa}       {\sin^2\!\thw}
\newcommand {\swsqeff}    {\sin^2\!\thweff}
\newcommand {\swsqefff}    {\sin^2\!\thwefff}
\newcommand {\swsqeffff}   {\sin^2\!\theta_{\rm{eff}}^{\rm {f}}}
\newcommand {\swsqeffl}    {\sin^2\!\theta_{\rm{eff}}^{\rm {lept}}}
\newcommand {\ee}         {\mathrm{e}^+\mathrm{e}^-}
\newcommand {\ppb}        {p\bar{p}}
\newcommand {\roots}      {\sqrt{s}}
\newcommand {\Rbb}        {{R_{\mathrm{b}}}}
\newcommand {\Rcc}        {{R_{\mathrm{c}}}}
\newcommand {\Afbzb}     {A^{0,\,{\rm b}}_{\rm {FB}}}
\newcommand {\Afbzc}     {A^{0,\,{\rm c}}_{\rm {FB}}}
\newcommand {\Ab}      {\rm{A_b}}
\newcommand {\Ac}      {\rm{A_c}}
\newcommand {\Ae}      {\rm{A_e}}
\newcommand {\eeww}       {\ee\rightarrow\mathrm{W^+W^-}}
\newcommand {\qqb}        {q\bar{q}}
\newcommand {\qqbp}       {q\bar{q}^{'}}
\newcommand {\bbb}        {b\bar{b}}
\newcommand {\ttb}        {t\bar{t}}

%%%end of pbr definitions

\title{Global Electroweak Fits and the Higgs Boson Mass} 

\author{Peter Renton}
\affiliation{Oxford University, Oxford, OX1 3RH, UK:   p.renton1@physics.ox.ac.uk}

\begin{abstract}
The current electroweak data and the constraints on the Higgs mass are
discussed. Within the context of the Standard Model the data prefer a
relatively light Higgs mass.

\end{abstract}

\maketitle

\thispagestyle{fancy}


\section{PRECISION ELECTROWEAK DATA} % Section title should be in all capitals.

This report contains an update on the values of the precision electroweak
properties and fits within the context of the Standard Model (SM), with respect 
to~\cite{ewwg2007}, where more details can be found. 
The $\ee$ data are from the ALEPH, DELPHI, L3 and OPAL experiments at LEP, 
and from the SLD experiment at SLAC. All the LEP1 results at the Z-pole 
are final~\cite{physrep}. The $\ppb$ data come from the CDF and
D0 experiments at the Tevatron, using integrated Run 2 luminosities of
up to 2.8 fb$^{-1}$.

The Z-lepton couplings (see~\cite{ewwg2007,physrep} for definitions and details) 
are extracted from 
the $\tau$ polarisation (A$_{e}$, A$_{\tau}$), the SLAC polarised 
electron asymmetry $\ALR$ (A$_{e}$) and the forward-backward asymmetries for 
leptons (A$_{\ell}$, $\ell$=e,$\mu,\tau$). The results are reasonably 
compatible with lepton universality and, assuming this, 
give $\Ae$ = 0.1501 $\pm$ 0.0016.
Within the context of the SM this favours a light Higgs mass.
The invisible width of the Z boson allows the number of light neutrinos 
to be extracted (assuming $\Gamma_\nu/\Gamma_l$ from the SM), 
and gives N$_{\nu}$ = 2.9841 $\pm$ 0.0083, which is 1.9 $\sigma$ below 3.

In addition to these results, which involve only the Z-lepton couplings, 
there are also results involving Z-quark couplings. There are six
such heavy-flavour quantities used; namely, the partial hadronic branching
ratios and pole forward-backward asymmetries for b and c quarks ($\Rbb$, $\Rcc$, 
$\Afbzb$ and $\Afbzc$) and the quantities $\Ab$ and $\Ac$, measured directly
by the SLC using a polarised electron beam.

There are six determinations of the effective weak mixing 
angle, giving an average value $\swsqeffl$=0.23153 $\pm$ 0.00016~\cite{ewwg2007}. 
There is a long-standing and {\it a posteriori} observation that the value obtained
from purely leptonic processes ($\swsqeffl$=0.23113 $\pm$ 0.00021) is some
3.2$\sigma$ different to that obtained using heavy quarks 
($\swsqeffl$=0.23222 $\pm$ 0.00027). This comes mostly
from the 3.2$\sigma$ difference between the SLD $\ALR$ and the $\Afbzb$ values.
The heavy-flavour results favour a rather heavy Higgs boson.
However, it is worth noting that the overall $\chisq$ probability for the
compatibility of all 6 measurements is reasonable (3.8$\%$).

The W boson is produced singly at the Tevatron (eg 
$u+\bar{d} \rightarrow W^{+}$). The leptonic decays W$\rightarrow \ell\nu$ 
(with $\ell = e,\mu$) are used to determine the W mass and width, using 
the transverse mass, p$_{T}^{\ell}$ or p$_{T}^{\nu}$. CDF have published a Run 2
measurement, using an integrated luminosity of $\simeq$ 0.2 fb$^{-1}$,
which gives $\MW$ = 80.413 $\pm$ 0.048 GeV; the single most precise 
experimental value. The Tevatron average has been recently 
updated (see \cite{tevewwg}, where details and references can be found), 
using a more consistent treatment of the Run 1 uncertainties on pdf's, 
electroweak corrections and the value of $\GW$ at which $\MW$ is determined.
This is important because the measured $\MW$ and $\GW$ values have a 
significant correlation. The SM value of $\GW$ has also been updated~\cite{pbr} to
$\GW$ = 2.093 $\pm$ 0.002 GeV, and the $\MW$ values are given for this value of $\GW$.
The updated Tevatron average is  $\MW$ = 80.432 $\pm$ 0.039 GeV.

At LEP2 the W bosons are pair-produced in $\eeww$. The individual results from
the four experiments are final and published, but the combination process 
is still preliminary. The statistical uncertainties from the
$\ell\nu\qqbp$ and $\qqbp\qqbp$ channels are similar. 
The Final State Interaction (FSI) uncertainties, which include non-perturbative 
colour reconnection (CR) and Bose-Einstein Correlation (BEC) effects in the
$\qqbp\qqbp$ final state, and which lead to `cross-talk' between the two W bosons, 
are still under study. 
At present a sizeable ($\simeq$ 36 MeV) common uncertainty is used, and this means that 
the $\qqbp\qqbp$
channel has only a 22\% weight in the combination with the $\ell\nu\qqbp$.
The preliminary LEP2 value is $\MW$ = 80.376 $\pm$ 0.033 GeV~\cite{ewwg2007}. 
This is uncorrelated with the Tevatron measurement, and combining all these 
gives $\MW$ = 80.399 $\pm$ 0.025 GeV. This value corresponds to a rather light
Higgs boson in the context of the SM.

The Tevatron W width, which includes a preliminary D0 and a published CDF value from Run 2,
has also been updated~\cite{tevewwg}, giving  $\GW$ = 2.050 $\pm$ 0.058 GeV.
For LEP2, the FSI uncertainty is still preliminary and the current preliminary LEP 
combined value is  $\GW$ = 2.196 $\pm$ 0.083 GeV. Together these give a 
revised World Average of $\GW$ = 2.098 $\pm$ 0.048 GeV, 
compatible with the SM expectation~\cite{pbr}.

In the SM the top quark decays mainly as t$\rightarrow$Wb.
The CDF and D0 Collaborations have continued to improve the precision on the
top-quark mass, using up to 2.8 fb$^{-1}$ of Run 2 data and a variety of methods. 
The most precise values come from the $\ttb \rightarrow \bbb\qqb\ell\nu$ final state.
The uncertainty in the jet energy scale (JES) is the largest potential systematic effect
and this is reduced by simultaneously fitting to $\Mt$ and a multiplicative JES factor, 
such that the $\qqb$ invariant mass is constrained to the well-known value of $\MW$.
The updated average value (see \cite{mt}, where details and references can be found) 
is $\Mt$ = 172.4 $\pm$ 0.7 (stat)  $\pm$ 1.0 (syst) GeV.
This gives a total uncertainty of 1.2 GeV, a relative precision of 0.7\%.
The experimental values of $\Mt$ extracted correspond to those used 
in the various Monte Carlo simulation programs. 
At present, any potential common systematic uncertainties 
associated with non-perturbative QCD effects (e.g. colour reconnection) are not
included.

\section{ELECTROWEAK FITS}

The SM parameters required for the electroweak fits 
are $\MZ$, G$_{F}$, $\alphamz$ and $\alphasmz$, 
(the electromagnetic and strong coupling constants at the scale $\MZ$), 
and the top-quark mass $\Mt$. Through loop diagrams, measurements
of the precision electroweak quantities are sensitive to $\Mt$ (quadratically) and, 
to the `unknown' in the SM, $\MH$ (logarithmically). 
The SM computations use the programs TOPAZ0 and 
ZFITTER (for more details see ~\cite{ewwg2007}). 
The latter program (version 6.42) incorporates the  
fermion 2-loop corrections to $\swsqeffl$ and full 2-loop and leading 3-loop 
corrections to $\MW$~\cite{loops}.

The value of $\alpha$ at the scale $\MZ$ requires the use of data on
$\ee\rightarrow$ hadrons at low energies and the use of perturbative QCD
at higher energies. The various estimations of $\alphamz$ differ in the extent 
to which perturbative QCD is used, as well as in the data sets used in the evaluation. The
quantity needed is the hadronic contribution from the 5 lightest
quarks $\Delta\alpha^{(5)}_{\rm{had}}$,
and the value used by the LEP EWWG~\cite{ewwg2007} is 
$\Delta\alpha^{(5)}_{\rm{had}}$($\MZ$) = 0.02758 $\pm$ 0.00035~\cite{BP2005}.
New data, since the publication of ~\cite{BP2005}, particularly preliminary data from BES,
could have a sizeable influence on both the central value and uncertainty.
So finalisation of these BES results could have an important influence on the
results of the electroweak fits. It is worth noting that the present uncertainty
on $\Delta\alpha^{(5)}_{\rm{had}}$($\MZ$) corresponds to $\delta\MH/\MH \simeq$ 20\%.

\begin{figure*}[ht]
\centering
\includegraphics[scale=0.35]{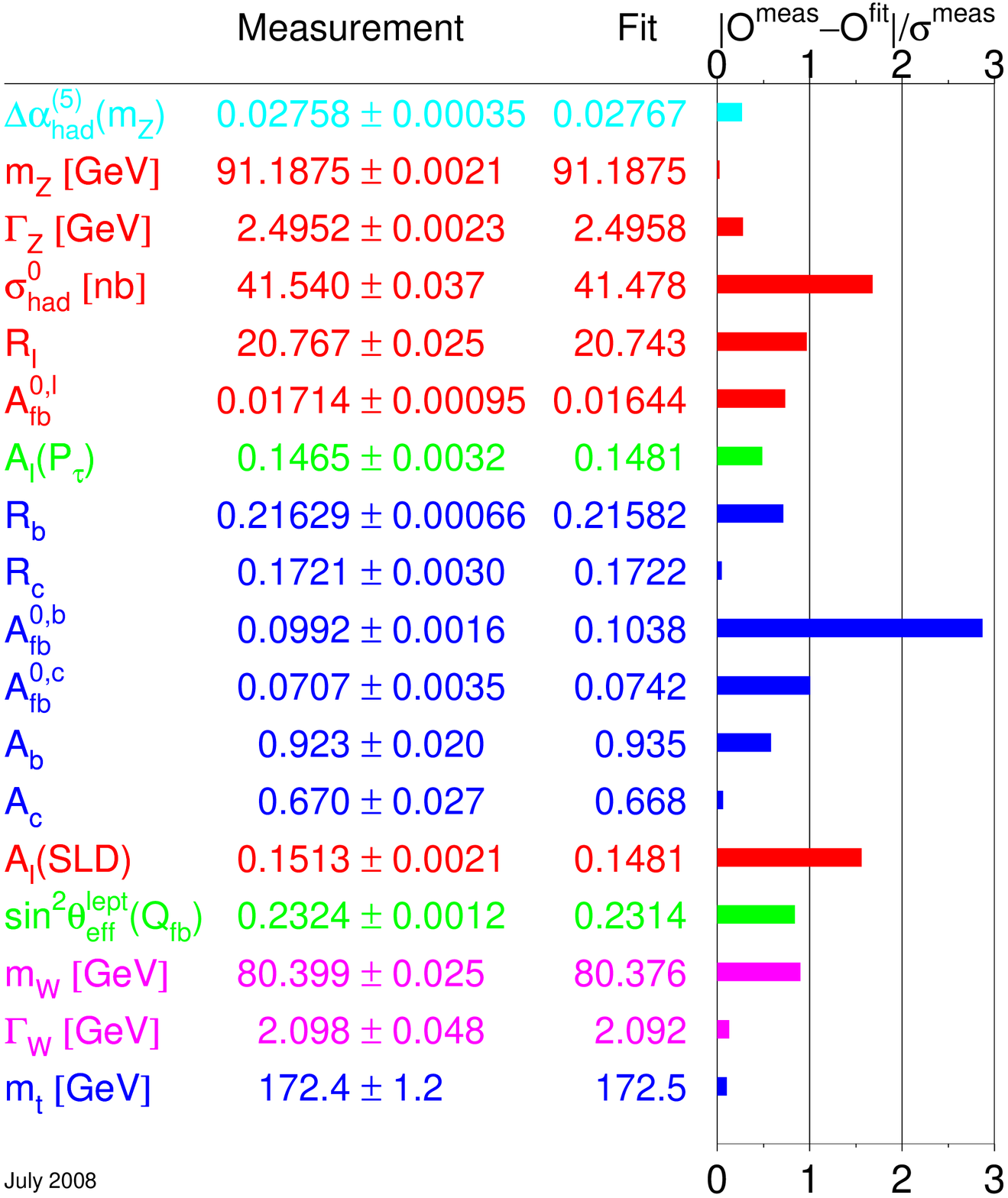}
\includegraphics[scale=0.45]{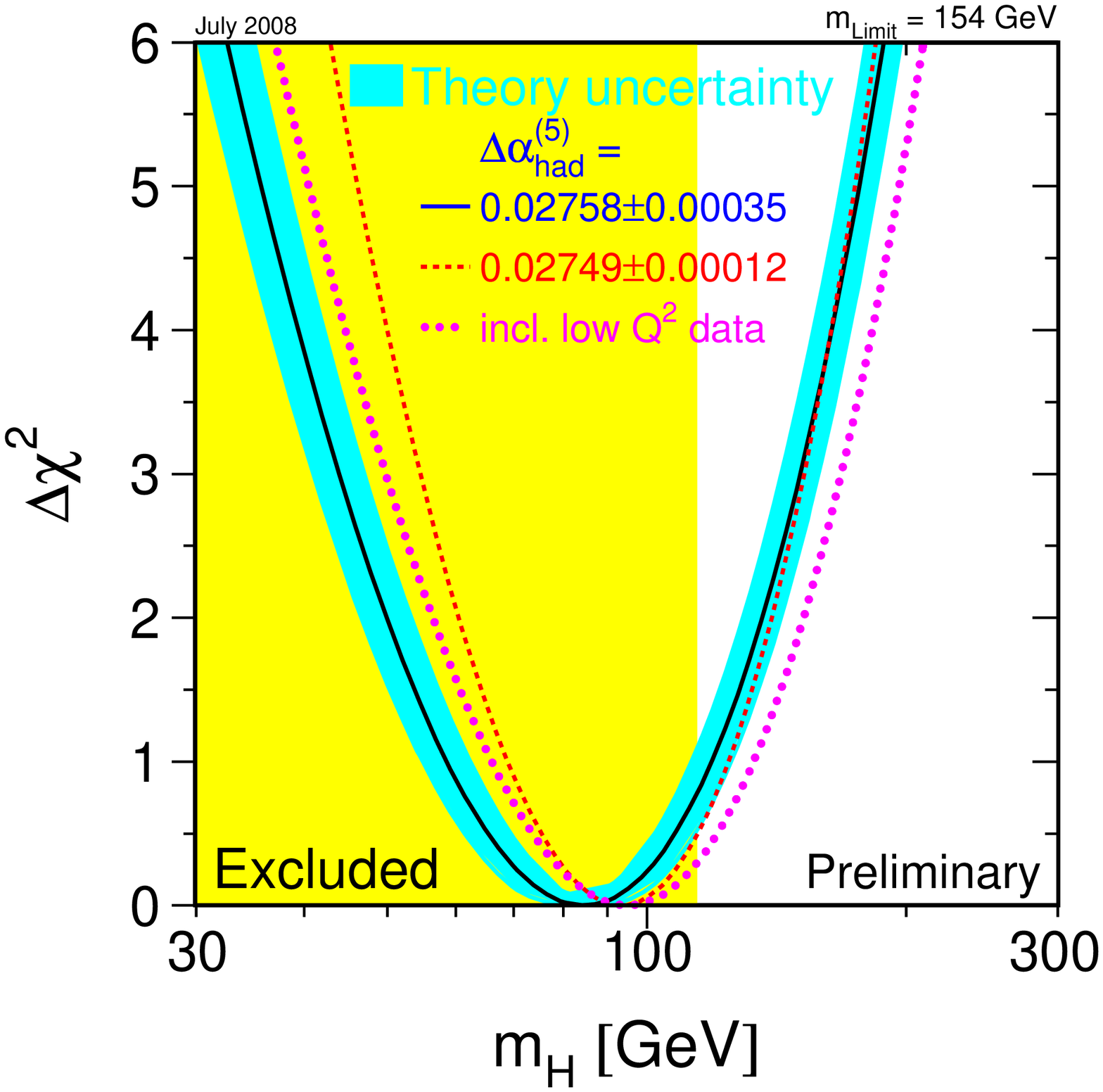}
\caption{Left: The measured and fitted values, together with the pull values.  
Right: The 'blue-band' plot showing the variation of $\chisq$ as a function
of $\MH$. The region excluded by the direct SM Higgs search at LEP2 is also shown.} 
\label{s08_higgs}
\end{figure*}

The 17 measurements used in the global SM electroweak fits, and the 
corresponding fitted values, 
are shown in fig.~\ref{s08_higgs}. The SM fit to these high Q$^{2}$ data
gives

\begin{center}
$m_{t}$  =  172.5 $\pm$ 1.2  GeV

$\alpha_{s}(\MZ)$  =  0.1185 $\pm$ 0.0026 

$m_{H}$  =  84 $^{ +34}_{ -26}$  GeV, 
\end{center}
with a  $\chisq$/ndf of 17.2/13; a probability of 19\%.

\begin{figure*}[ht]
\centering
\includegraphics[scale=0.45]{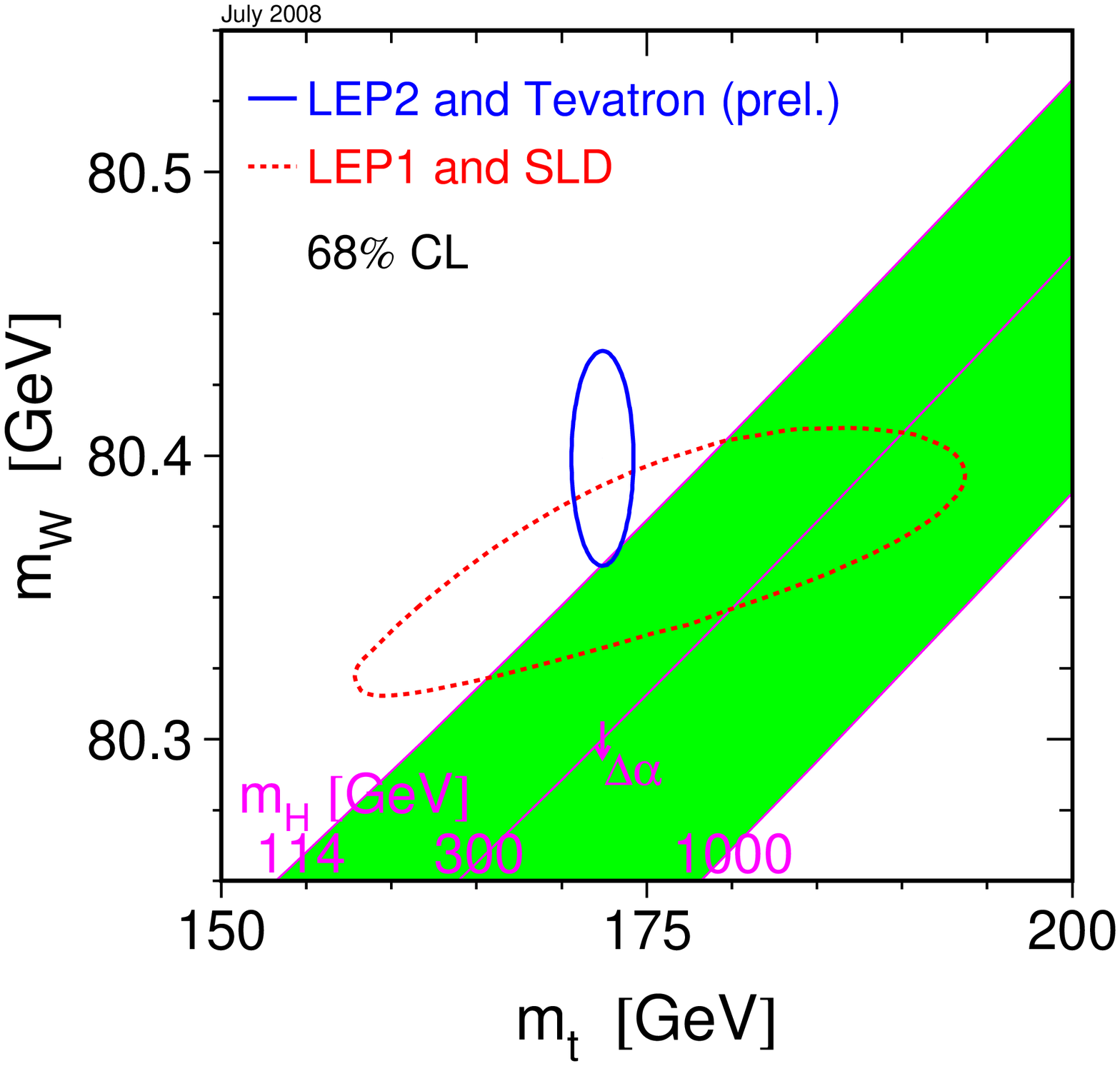}
\caption{68\% contours for $\Mt$ versus $\MW$ for both direct (solid line) and 
indirect measurments (dashed line). The corresponding Higgs mass values are also shown.}
\label{s08_mt_mw_contours}
\end{figure*}

 The variation of the fit $\chisq$, compared to the
minimum value, is shown in the `blue-band' plot of  fig.~\ref{s08_higgs},
as a function of $\MH$. Also shown is the direct SM Higgs search limit of 114 GeV from 
LEP2 searches. 
The one-sided 95\% upper limit is $\MH \leq$ 154 GeV. This includes the
theoretical uncertainty (blue-band), which is evaluated by considering 
the uncertainties in the 2-loop calculations~\cite{loops}.
If the more theory-driven value 
$\Delta\alpha^{(5)}_{\rm{had}}$($\MZ$) = 0.02749 $\pm$ 0.00012 is used, then the fitted
value of $\MH$ increases to 94 GeV. It is also interesting to note that there is 
now~\cite{ichep_higgs} a 95\% exclusion limit from the Tevatron 
at around $\MH \simeq $170 GeV.

Since the fits made in 2007~\cite{ewwg2007}, the main change is from the new top-quark 
mass (previous value $\Mt$ = 172.4 $\pm$ 1.8 GeV), resulting in an
increase in  $\MH$ of about 8 GeV with respect to \cite{ewwg2007}.

The quantities on which improved experimental precision can be expected in the
near future are $\Mt$, $\MW$, and $\Delta\alpha^{(5)}_{\rm{had}}$. The relative
current sensitivity to these quantities can be estimated as follows. If the
central value of $\Mt$, which is input to the fit, is changed 
by $\pm$ 1 $\sigma$ (i.e. $\pm$ 1.2 GeV), then the
corresponding shifts in the fitted values of $\MH$ are +9 GeV and -8 GeV
respectively. Similarly, for  $\pm$ 1 $\sigma$  changes in $\MW$ (i.e. $\pm$ 25 MeV),
the corresponding shifts in the fitted values of $\MH$ are -13 GeV and +17 GeV
respectively. For  $\pm$ 1 $\sigma$  changes 
in  $\Delta\alpha^{(5)}_{\rm{had}}$ (i.e. $\pm$ 0.00035), the corresponding shifts 
in the fitted values of $\MH$ are -15 GeV and +17 GeV respectively. So it can be seen that
improving the precision of  $\MW$ and  $\Delta\alpha^{(5)}_{\rm{had}}$ is
particularly important. 

Comparison of the direct versus indirect values of $\Mt$ and $\MW$ is a powerful test
of the SM; see fig.~\ref{s08_mt_mw_contours}. This method of presenting the 
electroweak data was first formulated in \cite{pbr_lp95}. The contours shown are for the
68\% cl. It can be seen that there is a reasonable degree of overlap 
and that both the direct and indirect data prefer a light Higgs mass.
Indeed, the region preferred by the data corresponds to that expected in MSSM SUSY models.

It is of interest to consider the effect of the future improved precision which 
can be expected from the Tevatron. 
Assuming that the uncertainty on $\Mt$ can be reduced from 1.2 
to 1.0 GeV, and that the uncertainty on the World Average value of $\MW$ can 
be reduced from 25 to 15 MeV, then, if the 
central values of all measured quantities remain the same, the fitted
Higgs mass would become
\begin{center}
$m_{H}$  =  71 $^{ +24}_{ -19}$  GeV, 
\end{center}
with a one-sided 95\% upper limit of 117 GeV. That is, this limit would not be far 
from the direct exclusion limit from LEP2. So the improved precision might lead
to the interesting situation where the results would be in conflict with the SM.

\section{SUMMARY}

The current electroweak data severely constrain the Standard Model
and prefer a relatively light Higgs boson mass.  Improvements in the accuracy
of the measurments used in the extraction of $\Delta\alpha^{(5)}_{\rm{had}}$ are important.
The improved precision on both $\Mt$ and $\MW$
expected from the Tevatron, and then the LHC, is easily awaited.

\begin{acknowledgments}
I would like to thank the LEP and Tevatron EWWGs, in particular Martin Grunewald, for
their help. 
\end{acknowledgments}

\end{document}